\documentclass[aps,prl,twocolumn,superscriptaddress]{revtex4}

\usepackage{graphicx}
\usepackage{color}
\usepackage{amsfonts}
\begin{document}
\title{Detecting the local transport properties and the dimensionality of transport of epitaxial graphene by a multi-point probe approach}

\author{Lucas Barreto}
\author{Edward Perkins}
\author{Jens Johannsen}
\author{S\o ren Ulstrup}
\affiliation{Department of Physics and Astronomy, Interdisciplinary Nanoscience Center, Aarhus University,
8000 Aarhus C, Denmark}
\author{Felix Fromm}
\author{Christian Raidel}
\affiliation{Lehrstuhl f{\"u}r Technische Physik, Universit{\"a}t Erlangen-N{\"u}rnberg, Erwin-Rommel-Strasse 1, D-91058 Erlangen, Germany}
\author{Thomas Seyller}
\affiliation{Institut f{\"u}r Physik, Technische Universit{\"a}t Chemnitz, Reichenhainer Strasse 70, D-09126 Chemnitz, Germany}
\author{Philip Hofmann}
\affiliation{Department of Physics and Astronomy, Interdisciplinary Nanoscience Center, Aarhus University,
8000 Aarhus C, Denmark}
\email[]{philip@phys.au.dk}

\date{\today}

\begin{abstract}
The electronic transport properties of epitaxial monolayer graphene (MLG) and hydrogen-intercalated quasi free-standing bilayer graphene (QFBLG) on SiC(0001) are investigated by micro multi-point probes. Using a probe with 12 contacts, we perform four-point probe measurements with the possibility to effectively vary the contact spacing over more than one order of magnitude, allowing us to establish that the transport is purely two-dimensional. Combined with the carrier density obtained by angle-resolved photoemission spectroscopy, we find the room temperature mobility of MLG to be $ (870\pm120)\;\mathrm{cm^2/Vs}$. The transport in QFBLG is also found to be two-dimensional with a mobility of  $ (1600\pm160)\;\mathrm{cm^2/Vs}$. 
\end{abstract}


\maketitle

The unusual electronic structure of graphene  \cite{Novoselov:2004,Novoselov:2005}  has many important consequences for the electronic transport properties of this material and these have been widely studied \cite{Das-Sarma:2011}. In most transport experiments, exfoliated graphene is placed on insulating SiO$_2$, so that the carrier density can be changed by electric field gating, and lithographic methods are used to fabricate a device out of this structure. For device applications, this approach has some disadvantages, most importantly the small size and ill-defined position of the flakes and the possibility of defect creation in the transfer process. 

An alternative to exfoliated graphene is the direct synthesis of epitaxial graphene on single-crystal transition metal surfaces \cite{Sutter:2008,Vazquez-de-Parga:2008} or on SiC \cite{Forbeaux:1998}. This produces high-quality, single-crystal graphene, covering the entire surface of the sample.  However, when a metal is chosen as a substrate for the synthesis, the system cannot be used for transport because the conduction through the bulk metal would be dominant \cite{Lizzit:2012}. In the case of SiC, a wide-gap semiconductor, graphene-dominated transport measurements are easily possible \cite{Berger:2006,Jobst:2010}. Again, this is done by using lithographic techniques to construct graphene-based devices. In the data analysis, it is usually assumed that the entire conduction proceeds through the graphene and the role of the substrate is negligible. This assumption is very likely to be valid but it has not been confirmed experimentally. 

In this paper, we present an alternative approach to transport studies. We perform four-point probe measurements on epitaxial graphene on SiC by using a monolithic probe that can be approached to any desired position on the sample without the need for the construction of a device, and we perform the measurements in ultra-high vacuum (UHV). By using a probe with not only four but twelve contacts, we are able to effectively vary the contact spacing to reveal that the electrical transport is in fact two-dimensional and thus graphene-dominated. The effective variation of the contact spacing and the possibility to probe many different locations on the sample also gives information about the homogeneity of the sample, something that is difficult to obtain in a conventional transport experiment. 

\begin{figure}[t]
\begin{center}
\includegraphics[width=1\columnwidth]{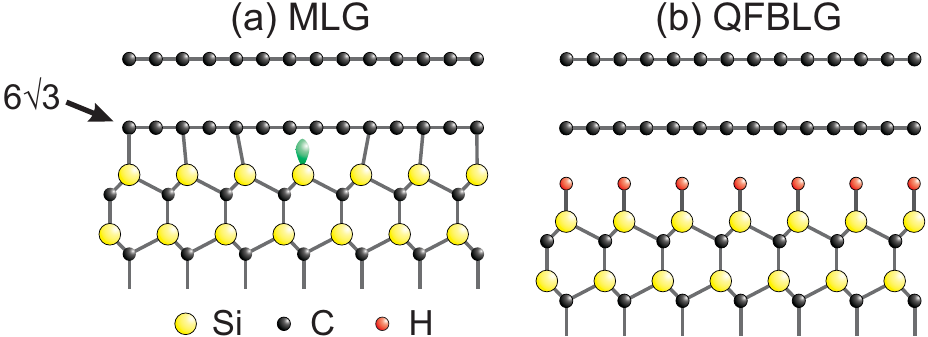}
\caption{\label{structures} (Color online) Schematic structures (side view) of the samples used in the present study. (a) Monolayer graphene (MLG) situated on the buffer layer ($6\sqrt{3}$). (b) Quasi free-standing bilayer graphene (QFBLG) on top of the hydrogen-saturated SiC surface. QFBLG is obtained by annealing MLG in molecular hydrogen. Drawing not to scale.}
\end{center}
\end{figure}

Two different types of graphene samples on SiC(0001) have been used. The first one, so-called monolayer graphene (MLG), is produced by the thermal decomposition of the Si-terminated face of SiC(0001). Heating this SiC face to sufficiently high temperatures results in the surface being terminated by a graphene-like carbon lattice that gives rise to a  $(6\sqrt{3} \times 6\sqrt{3})$R$30^{\circ}$ periodicity ($6\sqrt{3}$ in short). While the surface layer has the structure of graphene, it does not show graphene's characteristic electronic structure, due to the strong interaction with the underlying Si atoms. This graphene-like layer is called the buffer layer. Further heating of the surface results in the formation of a second layer with the graphene structure above the buffer layer as shown in Fig. \ref{structures}(a). This layer has the electronic properties of graphene and is termed MLG. The synthesis of the second type of samples used in this work, the so-called quasi free-standing  bilayer graphene (QFBLG) proceeds via hydrogen intercalation, starting with the MLG samples. Hydrogen intercalates below the buffer layer and binds to the underlying Si atoms. This leads to a structure of two decoupled graphene layers, the top layer and the previous buffer layer, and hence to QFBLG  \cite{Riedl:2009} which is depicted in Fig. \ref{structures}(b). The hydrogen-induced QFBLG formation is known to lead to a higher mobility than for MLG  \cite{Robinson:2011}.

More precisely, MLG and QFBLG were synthesized ontop of semi-insulating 6H-SiC(0001) substrates with a resistivity exceeding $10^{5}~\mathrm{\Omega cm}$ purchased from II-VI Inc. Details for the synthesis of  MLG and hydrogen intercalation are found in Refs. \onlinecite{Emtsev:2009,Ostler:2010} and \onlinecite{Speck:2010,Speck:2011}, respectively. In short, MLG was grown by annealing hydrogen-etched SiC(0001) samples in Ar at temperatures of around $1650^\circ \mathrm{C}$. The conversion of MLG into QFBLG was performed by annealing for one hour in 950~mbar ultra-pure molecular hydrogen at $850^\circ \mathrm{C}$. After confirmation of graphene coverage and buffer layer conversion by x-ray photoelectron spectroscopy, the samples were shipped from Erlangen to Aarhus. There, the samples were inserted into the UHV  chamber containing the multi-point probe setup (base pressure $\sim 5\times10^{-10}\, \mathrm{mbar}$), annealed to $\approx 200^\circ \mathrm{C}$ to clean them, and allowed to cool to room temperature prior to measurements.  Note that UHV conditions are required to obtain reproducible conductance results on graphene, most probably due to water adsorption \cite{Moser:2007,Klarskov:2011}. We have also performed angle-resolved photoemission (ARPES) measurements on the same samples, using the same \emph{in situ} cleaning by heating. These measurements were taken on the SGM-3  beamline at the synchrotron radiation source ASTRID \cite{Hoffmann:2004}.

\begin{figure*}[t]
\begin{center}
\includegraphics[width=2 \columnwidth]{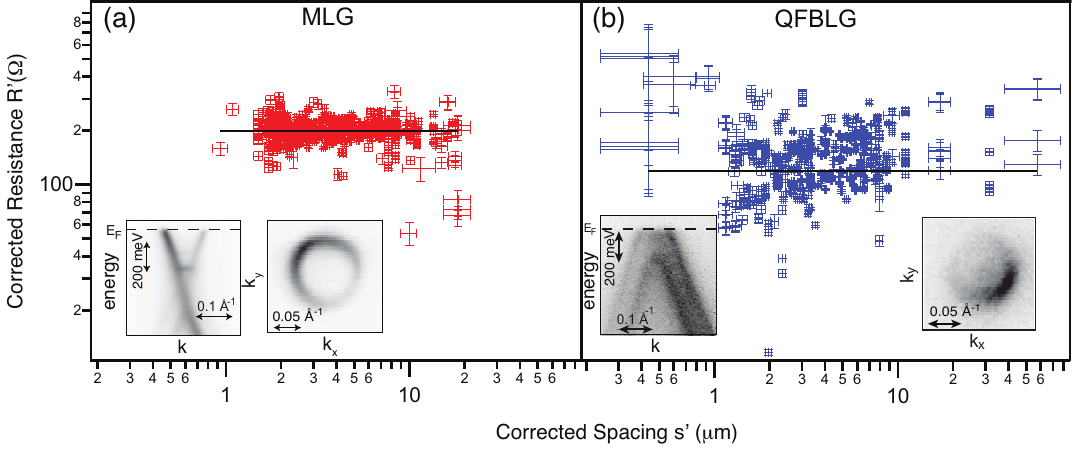}
\caption{(Color online) Corrected resistance $R'$ as a function of the corrected spacing $s'$ for (a) monolayer graphene on SiC(0001) (MLG) and (b) quasi free-standing bilayer graphene on SiC (QFBLG).  The black lines are best fits according to Ref. \onlinecite{Wells:2008c}, giving $\sigma_s=1.1\;\mathrm{mS}$ and $\sigma_s=1.8\;\mathrm{mS}$ respectively. The insets show the $\pi$-band dispersion near the $\bar{K}$ point of the Brillouin zone (left) and Fermi contour (right). \label{4pp_data}}
\end{center}
\end{figure*}

Lateral transport measurements are performed by placing a microscopic 12-point probe \cite{capres}  onto the sample surface using piezoelectric motors. For an equidistant collinear four-point probe, the expected measured four-point probe resistances (measured voltage between inner probes divided by current through outer probes) for a two-dimensional and a semi-infinite three-dimensional sample are $R_{2D}=\ln 2 / \pi \sigma_s$ and $R_{3D}=1/2\pi s \sigma_b$, respectively, where $\sigma_s$ is the sheet conductivity, $\sigma_b$ the bulk conductivity and $s$ the contact spacing. The resistance is defined as the voltage drop over the two inner contacts divided by the current through the outer contacts. In such an experiment, the dimensionality of the transport (2D vs 3D) can be determined by changing the contact distance  \cite{Shiraki:2001,Hofmann:2009}. This is not an option for a monolithic probe with a fixed contact spacing but we have recently shown that a multi-point probe can be used to emulate a variable contact spacing by choosing different combinations of contacts as current sources and voltage reference \cite{Wells:2008c}.

Actual resistance measurements are made by ramping the current through the sample while measuring the voltage drop, leading to an $I/V$ curve that gives the resistance via a linear fit.  The $I/V$ curves are routinely taken with the current ramped in both directions because this permits the easy identification of capacitance-induced artefacts. Then, instead of plotting the measured resistance as a function of contact spacing, we apply a geometric transformation, that allows us to plot a graph between the ``corrected resistance'' $R' = \chi_{2D} R$ and the ``corrected spacing'' $s'=s_{eff}/\chi_{2D}$, quantities with units of ohms and metres, respectively \cite{Wells:2008c,commmanes}. As in the case of a variable spacing, $R'$ is independent of $s'$ for 2D transport and inversely proportional to $s'$ for 3D transport. The most significant assumption made in this treatment is that the surface and bulk can be treated as parallel resistors.

Fig. \ref{4pp_data}(a) shows the acquired data for MLG. The plot contains 694 data points measured for different values of $s'$ at different positions on the sample. Each point  was averaged from at least 3 different measured $I/V$ curves. $R'$ shows no trend vs. $s'$, and fitting with a line confirms this, giving a result consistent with zero bulk conductivity. This shows that the current flows in a 2D fashion which agrees with our expectation that the graphene carries the current. The absence of any significant penetration into the substrate bulk - even at larger values of $s'$ - is consistent with the high resistivity of the underlying semi-insulating SiC substrate ($\rho \ge 10^{5}~\mathrm{\Omega cm}$). Note that the graphene growth does not alter the insulating properties of the substrates \cite{Emtsev:2009}.

ARPES data from the same sample is shown as an inset in Fig. \ref{4pp_data}(a). The band dispersion near the $\bar{K}$ point of the Brillouin zone shows graphene to be strongly $n$-doped with the Dirac point $\approx 360$~meV below the Fermi energy. The carrier concentration can be directly calculated from the Dirac point position and the known dispersion of the graphene $\pi$- and $\pi^{*}$-bands, giving a value of $(8.4\pm1.1)\times10^{12}\;\mathrm{electrons/cm^{2}}$, which is in good agreement  with previous work \cite{Bostwick:2007,Jobst:2010,Johannsen:2012}. When combining the carrier concentration with the measured sheet conductivity, we obtain a carrier mobility of $(870\pm120)\;\mathrm{cm^2/Vs}$. This value and even the degree of scattering in the data is in good agreement with the results of transport measurements on graphene Hall-bar devices fabricated from MLG \cite{Jobst:2010}.

Fig. \ref{4pp_data}(b) shows the corresponding data set of 571 points obtained from QFBLG. This sample also shows a two-dimensional behaviour, with a higher surface conductivity of $\sigma_s=1.8\;\mathrm{mS}$. The degree of scatter in the data points is somewhat larger than for MLG, something that could be related to slightly inhomogeneous hydrogen intercalation. ARPES shows a clear $p$-doping with the Dirac point $\approx 290$~meV above the Fermi energy. The $p$-doping is consistent with previous studies of QFBLG \cite{Riedl:2009,Robinson:2011,Tanabe:2012}, even though the absolute degree of doping is somewhat stronger here. We derive a carrier concentration of $(7.1\pm0.7 )\times10^{12}\;\mathrm{holes/cm^{2}}$.  By combining these results, we obtain a carrier mobility of $(1600\pm160)\;\mathrm{cm^2/Vs}$. The increase of mobility and also its absolute value is consistent with literature reports \cite{Tanabe2012,Robinson:2011}.

For both types of samples investigated here, we have thus demonstrated that the effective variation of the contact spacing permitted by a multi-point probe directly reveals a corrected resistance that is independent of the corrected contact spacing, i.e. the presence of two-dimensional transport.  This may not be surprising for graphene on SiC,  but it is in drastic contrast to a study of Bi(111), a semimetal surface with metallic surface states \cite{Hofmann:2006}, for which only the three dimensional bulk transport has been detectable \cite{Wells:2008c}. For many interesting materials, such as the topological insulators \cite{Hasan:2010}, there will be a competition between bulk and surface transport and the approach presented here should be able to single out the surface contribution. 

The conductivity and mobility for both graphene systems measured here in UHV are quite similar to earlier room temperature results obtained in air. This suggests that contamination-induced mobility reduction is not a significant factor for graphene at room temperature, unlike the situation in low temperature measurements \cite{Bolotin:2008}.

This work has been supported by the Carlsberg Foundation and Danish Council for Independent Research - Technology and Production Sciences. Work in Erlangen was supported by the European Union through the CRP CONCEPT GRAPHENE. We thank  Justin Wells and Lauge Gammelgaard for fruitful discussions and CAPRES A/S for technical support.


\end{document}